\documentclass[twocolumn]{aastex62}

\def\al{\alpha}

\def\ga{\gamma}
\def\de{\delta}

\def\th{\theta}

\def\ka{\kappa}

\def\Om{\Omega}

\shortauthors{Li \& Wang }

\begin{document}

\title{Energy release from magnetospheres deformed by gravitational waves}

\correspondingauthor{Huiquan Li}\email{lhq@ynao.ac.cn}

\author{Huiquan Li}
\affil{Yunnan Observatories, Chinese Academy of Sciences, \\
650216 Kunming, China}
\affil{Key Laboratory for the Structure and
Evolution of Celestial Objects,
\\ Chinese Academy of Sciences, 650216 Kunming, China}
\affil{Center for Astronomical Mega-Science, Chinese Academy of
Sciences, \\ 100012 Beijing, China}

\author{Jiancheng Wang}
\affil{Yunnan Observatories, Chinese Academy of Sciences, \\
650216 Kunming, China}
\affil{Key Laboratory for the Structure and
Evolution of Celestial Objects,
\\ Chinese Academy of Sciences, 650216 Kunming, China}
\affil{Center for Astronomical Mega-Science, Chinese Academy of
Sciences, \\ 100012 Beijing, China}

\begin{abstract}

In this work, we consider the possibility of energy release in
pulsar magnetospheres deformed by gravitational waves from nearby
sources. The strong electromagnetic fields in the magnetospheres may
release non-negligible energy despite the weakness of the
gravitational wave. When the background spacetime is perturbed due
to the passage of a gravitational wave, the original force-free
state of the inner magnetosphere will be slightly violated. The
plasma-filled magnetosphere tends to evolve into new force-free
states as the spacetime varies with time. During this process, a
small portion of the electromagnetic energy stored in the
magnetosphere will be released to the acceleration of charged
particles along the magnetic field lines. When the pulsar is close
enough to the gravitational wave source (e.g., $\sim10^{-2}$ pc to
the gravitational wave sources observed recently), the resulting
energy loss rate is comparable with the radio luminosity of the
pulsar. It is also noticed that, under very stringent conditions
(for magnetars with much shorter distance to the sources), the
released energy can reach the typical energy observed from fast
radio bursts (FRBs).

\end{abstract}


\section{Introduction}
\label{sec:introduction}
%%%%%%%%%%%%%%%%%%%%%%%%%%%%%%%%%%%%%%%%%%%%%%%%%%%%%%%%%%%%%%%%%%%%%%%%%%%%

Magnetospheres on compact objects act as mediators through which
rotational energy are extracted from the objects to accelerate
charged particles and produce multi-band emission. Well within the
light cylinder (LC), the plasma-filled magnetosphere on a pulsar is
assumed to be force-free at large scales
\citep{1969ApJ...157..869G}, except in some small gaps where the
component of the electric field parallel to the magnetic field line
is not sufficiently screened \cite{1982RvMP...54....1M}. In the
force-freeness violated gaps, charged particles will be accelerated
by the parallel electric fields along magnetic field lines, which
will produce copious screening charges via the pair cascade
processes
\citep{1975ApJ...196...51R,1979ApJ...231..854A,1998ApJ...508..328H,
2000ApJ...532.1150Z,2001ApJ...560..871H,2017JApA...38...52M} to keep
the magnetosphere force-free.

In this work, we consider the dynamical process in the force-free
magnetosphere when its background spacetime varies due to the
passage of a gravitational wave (GW) from a nearby source. Though
the GWs are quite weak away from the sources, the effects of GWs on
magnetospheres may be compensated by the fact that the magnetic
fields on neutron stars are very strong, even reaching $10^{15}$
Gauss for magnetars. On a time-varying spacetime, the force-free
condition as well as Maxwell's equations for the original
magnetosphere will be slightly violated. The plasma-filled
magnetosphere tends to evolve into new force-free states. During the
process, part of the energy stored in the magnetosphere is triggered
to be released by the GW, via various mechanisms.

Based on some simple analysis, we estimate the energy released
during the process. We show that the energy is comparable to or even
larger than that of the pulsar low energy emission as the pulsar is
close enough to the source. This may provide an alternative
detecting signal for the GW, which is in spirit similar to the
method discussed in \citet{2018arXiv180400453D}. Even when the GW
sources are far away from us, we can detect them from the
observation of the timing and emission variability of a pulsar
located closed to the GW source.

This situation may exist in globular clusters with dense stars. In
the clusters, the distance between stars is short and is typically 1
light year (even shorter near their cores). In particular,
hierarchical triple-star systems can form frequently via
binary-binary encounters in the globular clusters
\citep{2016ApJ...816...65A}. Our discussion here is mostly relevant
to a hierarchical triple system in which a pulsar obits around a
close inner binary. As the binary is in the final stage of merger,
GW will be radiated and the pulsar magnetosphere will be perturbed.
Actually, a triple system involving a pulsar has been identified by
\citet{2014Natur.505..520R} (though the pulsar is in the inner
binary). Interestingly, it is long realised that, in a triple
system, the formation
\citep{2010ApJ...717..948I,2013MNRAS.431.2155N,2016ApJ...822L..24N}
and merger
\citep{2000ApJ...528L..17P,2016ApJ...816...65A,2017ApJ...836...39S}
of binaries can be enhanced via interactions with the outer
companion.

The pulsars, especially magnetars, are usually thought to be related
to the origin of the mysterious fast radio bursts (FRBs)
\citep{2007Sci...318..777L,2013Sci...341...53T,2016Natur.531..202S}
in many models proposed so far, though they encounter various
problems \citep{2018arXiv180603628P}. In some of the models, FRBs
arise from energy release and emission processes in pulsar
magnetospheres disturbed by eternal materials or sources
\citep{2014A&A...569A..86M,2015ApJ...809...24G,2016ApJ...822L...7W,
2017ApJ...836L..32Z,2017A&A...598A..88Z}. Here, we examine the
possibility of interpreting FRBs as energy release phenomena from
pulsar magnetospheres perturbed by GWs by comparing the energy
scales. The model has an attractive feature that the pulsar system
will not be contaminated, with nothing left behind, after the GW
passes through the magnetosphere.

%%%%%%%%%%%%%%%%%%%%%%%%%%%%%%%%%%%%%%%%%%%%%%%%%%%%%%%%%%%%%%%%%%%%%%%%%%%%
\section{Electrodynamics in the deformed magnetosphere}
\label{sec:electrodynamics}
%%%%%%%%%%%%%%%%%%%%%%%%%%%%%%%%%%%%%%%%%%%%%%%%%%%%%%%%%%%%%%%%%%%%%%%%%%%%

The plasma-filled magnetospheres of pulsars are usually assumed to
be force-free, with the inertia of plasma ignored:
\begin{equation}
 J^\mu F_{\mu\nu}=0,
\end{equation}
where $J^\mu$ is the 4-current. The spatial components of the
equation reads:
$\rho_e\textbf{E}+(1/c)\textbf{j}\times\textbf{B}=0$, i.e., the
Lorentz force on charges vanishes. This means that the electric
field must be perpendicular to the magnetic field everywhere in the
force-free magnetosphere:
\begin{equation}
 \textbf{E}\cdot\textbf{B}=0.
\end{equation}
The charge and current densities are given by Maxwell's equations:
\begin{equation}\label{e:Maxwellseq}
 \nabla\cdot\textbf{E}=4\pi\rho_e,
\textrm{ }\textrm{ }\textrm{ }
c\nabla\times\textbf{B}-\mathbf{\dot{E}}=4\pi\textbf{j}
\end{equation}

The associated spacetime metric to the above equations is generally
$g_{\mu\nu}$. For pulsar magnetospheres, we simply adopt flat
spacetime with $g_{\mu\nu}=\eta_{\mu\nu}$, as usually done in most
simplified pulsar models \footnote{Though the general relativistic
effects should be non-negligible near the neutron star
\citep{1992MNRAS.255...61M,1998ApJ...508..328H,2001MNRAS.322..723R},
this will make no significant difference in the results of our
discussion here since we are interested in the perturbed dynamics of
the magnetosphere caused by the GW, instead of the unperturbed part.
}. In the stationary and axisymmetric case, the electric field
induced by the rotation of the magnetic field is explicitly
expressed as
\begin{equation}\label{e:indE}
 \textbf{E}=-\frac{\textbf{v}}{c}\times\textbf{B},
\end{equation}
where $\textbf{v}=\bf{\Omega}\times\textbf{r}+\ka\textbf{B}$. We
focus on the inner region well within the light cylinder (LC) where
the electromagnetic fields are strong. In this region, the current
parallel to the magnetic field lines is negligible and the
electromagnetic fields corotates with the star, i.e., the fields are
purely poloidal with no toroidal component.

The energy density of the magnetosphere is electromagnetically
dominated. It is
\begin{equation}
 \varepsilon=\varepsilon_E+\varepsilon_B,
\end{equation}
where
\begin{equation}
 \varepsilon_E=\frac{E^2}{8\pi}, \textrm{ }\textrm{ }\textrm{ }
\varepsilon_B=\frac{B^2}{8\pi}.
\end{equation}

Now let us consider the case in which a GW passes through the
magnetosphere system so that the background spacetime changes with
time, with the metric becoming
$g_{\mu\nu}=\eta_{\mu\nu}+h_{\mu\nu}$. It is easy to infer that
Maxwell's equations and the force-free equation in flat spacetime
are generally not satisfied in the varying spacetime. The system is
now time dependent and tries to evolve to restore new balanced
force-free states. The geometry of the field lines is deformed, and
the charge and current densities will redistribute in response to
the variability of the spacetime. Since the magnetosphere is
plasma-filled, the magnetosphere can quickly evolve into new
force-free states as the spacetime varies.

As the spacetime is perturbed, the energy density of the
electromagnetic system should change accordingly. The general form
of the change can be given by the energy-momentum tensor and is
spacetime dependent. Here, we just make an estimation of the scale
by doing some simple analysis. Let us consider the magnetosphere
penetrated by a GW whose propagation direction is along the $z$
coordinate and is perpendicular to the rotation axis of the
magnetosphere. Under the transverse-traceless (TT) gauge, the
quadrupole GW is decomposed into two polarisation directions along
the $x,y$ coordinates that are perpendicular to $z$. Accordingly,
the poloidal electromagnetic fields can also be decomposed in the
two directions. Then, in the linear order, the energy densities vary
as
\begin{equation}
 \de\varepsilon_E=\frac{1}{8\pi}[h_+(E_x^2-E_y^2)+2h_\times E_xE_y],
\end{equation}
\begin{equation}
 \de\varepsilon_B=\frac{1}{8\pi}[h_+(B_y^2-B_x^2)-2h_\times B_xB_y].
\end{equation}
Here, we ignore the rotational effect since we simply take the
varying energy densities as being on a similar scale for all slices.
Moreover, we are more interested in pulsars with strong magnetic
fields, which usually have long periods (e.g., longer than $\sim$
0.1 s), much longer than those of GWs from binary mergers (e.g.,
$\sim$ milliseconds for the observed GW events so far).

The axisymmetric magnetosphere can be viewed as being composed of
identical slices along the toroidal direction parameterized by the
toroidal angle $\phi$. On the two slices that are perpendicular to
the GW propagation direction $z$ (set to be at $\phi=0,\pi$), the
component electromagnetic fields satisfy $E^2=E_x^2+E_y^2$ and
$B^2=B_x^2+B_y^2$. When $E_x\sim E_y\sim E/\sqrt{2}$,
$\de\varepsilon_E\sim h_\times \varepsilon_E$, and when $E_x\sim E$
or $E_y\sim E$, $\de\varepsilon_E\sim \pm h_+ \varepsilon_E$. The
situation is similar for $\de\varepsilon_B$. So we can simply take
\begin{equation}\label{e:appenden}
 \de\varepsilon=\de\varepsilon_E+\de\varepsilon_B\sim h(\varepsilon_E
+\varepsilon_B).
\end{equation}
On other slices at $\phi\neq0,\pi$, the perturbed energy densities
are somewhat different, but should be of the similar scales.

These varying energy densities are those observed by asymptotic
observers. They are not necessarily dissipated or released during
the GW perturbation process. In what follows, we show that part of
the varying energy can indeed be transformed to the plasma in the
magnetosphere.

%%%%%%%%%%%%%%%%%%%%%%%%%%%%%%%%%%%%%%%%%%%%%%%%%%%%%%%%%%%%%%%%%%%%%%%%%%%%
\section{The energy release mechanisms}
\label{sec:mechanisms}
%%%%%%%%%%%%%%%%%%%%%%%%%%%%%%%%%%%%%%%%%%%%%%%%%%%%%%%%%%%%%%%%%%%%%%%%%%%%

In this section, we investigate the release processes of the
electromagnetic fields on the slices at $\phi=0,\pi$. It is
straightforward to extend the discussion to the cases on other
slices.

It is known that the relative distance change induced by the GW is
of the scale: $\delta L/L=h/2$. For the centered dipole
magnetosphere, we can estimate that the angular change $\al$ of the
field lines with respect to the center of the star should be $\sim
\delta L/L$, i.e.,
\begin{equation}
 \al\sim \frac{h}{2}.
\end{equation}
In the inner region of the magnetosphere, we can simply think that
this angular change is comparable with the change of the poloidal
angle $\th$ of the magnetic field lines: $\al\sim\de\th$.

\subsection{The emergence of parallel electric fields}

In the force-free magnetosphere, the electric and magnetic fields
are originally perpendicular to each other. As the spacetime is
perturbed by the GW, the relative angle between the electric and
magnetic field lines is $2\al$ different from $\pi/2$. Thus, there
emerges a non-vanishing component of the electric field along the
magnetic field line, which is of the scale:
\begin{equation}
 E_{\|}=E\sin(2\al)\sim hE.
\end{equation}

The charges on the magnetic field lines should be accelerated by
this parallel electric field. This point is easily seen in the local
flat frame. In a general spacetime that is relevant here, we can
always find a local Minkowski spacetime in the neighborhood of any
point. In the local frame, an observer at the point can see that the
angle between the electric and magnetic fields around will be
changed when the background spacetime is perturbed. Applying the
laws in flat spacetime, it is easy to know that, in the local flat
frame, the charges at the point should be accelerated by the
electric field parallel to the magnetic field line.

The acceleration process should be efficient under particular
conditions. Let us consider the ideal case in which a test electron
is accelerated by this emerging parallel electric field. For
$h\sim10^{-10}$ and $E\sim10^{11}$ V/m, the electron can be
accelerated to the relativistic speed with the Lorentz factor
$\ga=eE_{\|}\delta t/m_e c\sim5.8$ in a timescale $\de t\sim10^{-3}$
s. For an electron-positron pair, this is sufficient to cause them
to be separated by a distance of $\sim100$ km.

Of course, this is not the case in a real magnetosphere because the
magnetosphere is filled with plasma, with a density that is even
much more larger than the one needed to sustain the force-freeness
of the magnetosphere
\citep{1975ApJ...196...51R,2017JApA...38...52M}. The emerging
$E_{\|}$ (very small compared with $E$) will be quickly screened by
the collection of particle pairs on the conducting magnetic field
lines. During the process, its energy will be all transferred to the
acceleration of the charged particles. The released energy is
\begin{equation}
 (\mathcal{L}_E)_{\|}=\frac{E_{\|}^2}{8\pi}\sim h^2
\varepsilon_E.
\end{equation}

\subsection{The emergence of toroidal magnetic fields}

As seen from an observer corotating with the magnetic field line,
its projected distance to the rotation axis varies as the GW passes.
From Eq.\ (\ref{e:indE}), this means that the electric field induced
from the rotation of the magnetic field should vary accordingly
under the force-free condition. In the flat spacetime, the electric
field is given by the equation (\ref{e:indE}): $E=x\Om B$, with
$x=r\sin\th$. When the spacetime is perturbed, the distance $x$
changes by $\de x$. So the induced electric field satisfying the
force-free condition is $E'\equiv E-\de E=(x-\de x)\Om B$. With $\de
x/x\sim h/2$, we have
\begin{equation}
 \de E\sim\frac{1}{2}hE.
\end{equation}
So the energy density change from the variability of the
perpendicular electric field is
\begin{equation}\label{e:dissEper}
 (\mathcal{L}_E)_\bot=\frac{1}{8\pi}(E^2-{E'}^2)
\sim\frac{1}{4}h(4-h)\varepsilon_E.
\end{equation}
This is comparable with the variable energy density
(\ref{e:appenden}) of the electric field in linear order.

We now determine where the energy goes. When a magnetic field line
is shifted toward the rotation axis ($\de x>0$) at a moment, it will
still keep rotating with the original angular velocity, which is
larger than the angular velocity of corotation on the new ``orbit''.
Thus, a toroidal component $B^\phi$ of the magnetic field appears
due to the differential rotation velocities. With the emergence of
$B^\phi$, charged particles will be accelerated along the magnetic
field line and a poloidal current will be excited, as given by
Maxwell's equations (\ref{e:Maxwellseq}): $c\nabla_r\times
B^\phi-\dot{E}^\th=4\pi j^\th$ and $c\nabla_\th\times
B^\phi-\dot{E}^r=4\pi j^r$. This process proceeds until the extra
magnetic field energy is exhausted and the field line corotates with
the star again on the new orbit. The energy released can be measured
by the difference of the energy densities (\ref{e:dissEper}) of the
electric field induced by the magnetic field rotating on the two
``orbits".

Similarly, when the magnetic field line is shifted away from the
rotation axis ($\de x<0$), it will also keep rotating with the
original velocity, which is lower than that of corotation on the new
orbit. A toroidal magnetic field also appears. The poloidal current
in the opposite direction is excited. But, in this process, the
energy density (\ref{e:dissEper}) is negative, which means that the
energy is compensated and extracted from the star.

\subsection{The vibrating magnetic field lines}

As the wave passes through the magnetosphere, the magnetic field
line is rotated with a small angle $\al$ by the tidal force, which
can induce a transient electric field during the rotating process.
On the slices at $\phi=0,\pi$, this transient electric field is
along the toroidal direction. Its value can be estimated from Eq.\
(\ref{e:indE}):
\begin{equation}
 E_T\sim-\frac{1}{c}\frac{r\al}{\de t}B.
\end{equation}
Here $r\al/\de t$ is the average vibrating velocity of the magnetic
field line, with the time interval $\de t$ to be a quarter of the
period of the GW.

This electric field is induced when the magnetic field line is
vibrating. It disappears when the magnetic field line halts at the
rotated angle $\al$. So the energy of this transient electric field
is completely released. The released energy is converted to the
acceleration of charges along the $B^\phi$ component mentioned in
the previous subsection. The amount of the released energy divided
by volume is
\begin{equation}
 \mathcal{L}_B=\frac{E_T^2}{8\pi}
\sim\frac{h^2r^2}{4c^2\de t^2}\varepsilon_B.
\end{equation}
A similar process of excitation of electric field by varying
magnetic field caused by star oscillation has been discussed in
\citet{2004MNRAS.352.1161R}.

%%%%%%%%%%%%%%%%%%%%%%%%%%%%%%%%%%%%%%%%%%%%%%%%%%%%%%%%%%%%%%%%%%%%%%%%%%%%
\section{The total released energy}
\label{sec:totalenergy}
%%%%%%%%%%%%%%%%%%%%%%%%%%%%%%%%%%%%%%%%%%%%%%%%%%%%%%%%%%%%%%%%%%%%%%%%%%%%

From the above analysis, the energy loss contains two components.
The varying electric field energy in leading order is almost all
released: $\mathcal{L}_E=(\mathcal{L}_E)_\bot+(\mathcal{L}_E)_{\|}
%\sim\de\varepsilon_E
\sim h\varepsilon_E$. But only a small fraction of the varying
magnetic field energy is released, which is of the order $h^2$. The
above analysis is implemented on the slices of the magnetosphere
that are perpendicular to the propagation direction of the GW. For
other slices away from $\phi=0,\pi$, we have similar energy release
processes. On these slices, the GW can induce a toroidal component
of the magnetic field, which can excite a current along the poloidal
magnetic field lines. The fast vibrating magnetic field lines can
also release some of its energy. Here, we simply take the results in
the previous section as those that apply generally in all regions in
the inner magnetosphere.

As usual, we assume a dipole structure of the magnetosphere within
the LC. Then, the fields depending on the radial distance are:
\begin{equation}
 B=B_0\left(\frac{r_0}{r}\right)^3,
\textrm{ }\textrm{ }\textrm{ } E=E_0\left(\frac{r_0}{r}\right)^2.
\end{equation}
The magnetic field at the surface of the star is
$B_0=6.4\times10^{19}\sqrt{P\dot{P}}$ G
\citep{1996ApJ...464..306U,2015MNRAS.447.2631V,2015MNRAS.449.3755V},
where $P$ is the period of the pulsar. At the the LC $r=R_{LC}$, the
electric field is equal to the magnetic field $B=E$, which
determines $E_0=(r_0/R_{LC})B_0=(r_0\Om/c)B_0$. The radius of the
light cylinder is $R_{LC}=cP/2\pi=4.8\times10^9P$ cm.

By integrating over the inner magnetosphere region, we can get the
total released energy in the deformed magnetosphere:
\begin{equation}
 L_E=1.4\times10^{25}\left(\frac{h}{10^{-10}}\right)
\left(\frac{10^{7}\textrm{ y}}{\tau}\right)\left(\frac{r_0}{10^{6}
\textrm{ cm}}\right)^5 \textrm{ erg},
\end{equation}
\begin{eqnarray}
 L_B=1.4\times10^{18}\left(\frac{h}{10^{-10}}\right)^2
\left(\frac{10^{-3}\textrm{ s}}{\de t}\right)^2 \nonumber
\\ \left(\frac{B_0}{10^{12}\textrm{ G}}\right)^2
\left(\frac{r_0}{10^{6}\textrm{ cm}}\right)^5 \textrm{ erg}.
\end{eqnarray}
where $\tau=P/(2\dot{P})$. The relative strain $h=10^{-10}$
corresponds to that at a distance $d=4.1\times10^{-3}$ pc to
GW150914 \citep{2016PhRvL.116f1102A} or that at $d=4\times10^{-2}$
pc to GW170817 \citep{2017PhRvL.119p1101A}. Of course, the GW
sources are not necessarily GW bursts. They may be persistent
sources, like binary systems, only if the pulsar is close enough to
the source.

This amount of energy is released within a timescale of a quarter of
the GW period. For the above GW events, the timescale is $\de t\sim
10^{-3}$ s (the frequency is $\nu\sim250$ Hz). So the energy release
rate is $1.4\times10^{28}$ erg/s, which is almost the radio
luminosity from a pulsar. This featured variability of radiation
should be detectable and discriminable, which may provide signals
indicating that a GW is passing through a pulsar magnetosphere.

We also notice that stringent conditions are needed to account for
the energy observed from FRBs, whose characteristic energy is as
high as $10^{39}$ erg at 1 Gpc distance
\citep{2017Natur.541...58C,2018arXiv180603628P}. For example, $L_B$
can reach the energy $10^{38}$ erg when $h\sim 10^{-4}$, $\de t\sim
10^{-4}$ s and $B_0\sim 5\times 10^{15}$ G. This requires a magnetar
magnetosphere perturbed by a GW with the frequency $\nu=500$ Hz and
with the strain corresponding to that at $d\sim10^{-3}$ A to
GW150914. But if FRBs are at the distance of 1 Mpc, the FRB energy
is of $\sim 10^{33}$ erg. The conditions for the energy are quite
relieved: $h\sim10^{-6}$, $\tau\sim 10^3$ y for $L_E$ and
$h\sim10^{-6}$, $\nu\sim250$ Hz, $B_0\sim3\times10^{15}$ G for
$L_B$.

Of course, here we have only done some estimation of the energy
scales with some simple analysis. The details of the release process
need more accurate study in more realistic situations, in
particular, with the aid of numerical tools. There may exist
improvement and enhancement on the estimated energy that can relieve
the above strict conditions. For example, the charges should be more
effectively accelerated closer to the star surface because the
electromagnetic fields are stronger in inner regions. The charged
particles will attain higher velocities and can catch up to those
accelerated in outer regions that have lower velocities. So the
charged particles accumulated in the emission region, which makes it
look like there is more energy released from the magnetosphere.
Moreover, we may also not ignore the effects of the GW on the polar
gap. The polar gap plays crucial roles in the pulsar activities,
sourcing the plasma and the energy for radiation in the
magnetosphere, though it is quite small compared to the whole
magnetosphere. As shown in
\citet{2012A&A...540A.126Z,2015ApJ...799..152L}, the oscillating
modes of the star can drive the oscillation of the vacuum electric
field in the gap, bringing prominent changes in observational
features. So the perturbations of the GWs on the polar gap could
also cause similar effects on the gap electric field, which is much
stronger than the emergent electric fields discussed here. This may
cause extra energy release, awaiting for further examination.

\acknowledgments

This work is supported by the Yunnan Natural Science Foundation
2017FB005.

%\bibliographystyle{aasjournal}
%\bibliography{b}

\end{document}